\documentclass[10pt,letterpaper]{article}
\usepackage{opex3}
\usepackage{color}
\usepackage{epstopdf}

\begin{document}
\newcommand{\LN}{LiNbO$_3$ }
\newcommand{\Q}{$Q$ }

\title{Free-standing high quality factor thin-film lithium niobate micro-photonic disk resonators}

\author{Renyuan Wang$^*$ and Sunil A. Bhave}

\address{School of Electrical and Computer Engineering, Cornell University, Ithaca, NY, 14853}

\email{$^*$rw364@cornell.edu} 



\begin{abstract}
Lithium Niobate (LN or just “niobate”) thin-film micro-photonic resonators have promising prospects in many applications including high efficiency electro-optic modulators, optomechanics and nonlinear optics. This paper presents free-standing thin-film lithium niobate photonic resonators on a silicon platform using MEMS fabrication technology. We fabricated a 35um radius niobate disk resonator that exhibits high intrinsic optical quality factor (Q) of 484,000. Exploiting the optomechanical interaction from the released free-standing structure and high optical Q, we were able to demonstrate acousto-optic modulation from these devices by exciting a 56MHz radial breathing mechanical mode (mechanical Q of 2700) using a probe. 
\end{abstract}

\ocis{(000.0000) General.} 


\section{Introduction}

The lack of inversion symmetry in the Lithium Niobate (LN or niobate) crystal exhibits itself via its characteristic strong piezoelectric effect, linear electro-optic effect, pyroelectric effect and strong second order optical nonlinearity \cite{LNProperties}. Leveraging the strong electro-optic effect and optical nonlinearity, the optics industry uses bulk \LN for manufacturing photonic modulators \cite{LNMod} and optical frequency doublers \cite{PPLNComb}. In recent years, \LN thin-film photonics have attracted a lot of attention \cite{GunterLN, FemtoSTLN} for advanced photonic applications, as the micro-fabricated photonic structures can enable enhanced light/matter interaction from the highly confined optical mode and long interaction time. These can lead to reduced threshold excitation power, enhanced nonlinear optical effect, and agile electrically tunable photonic devices.

Microring resonators and microdisk resonators are fundamental building blocks for many micro-photonic devices including filters \cite{OpFilter}, lasers \cite{LaserRing}, and modulators \cite{OpMod}. A high \Q thin-film \LN resonator has the potential to enable many novel and high performance devices. While an-isotropic etching defined thin-film \LN photonic structures are desired for applications such as photonic crystals and resonators, the \Q demonstrated to date from such devices is generally limited to few thousand, while the quality factor from bulk \LN disk resonators can exceed 10$^8$ \cite{LNBulk}. The key limiting factor of the quality factor is the optical scattering loss from the rough side wall profile from the etching process. Surface smoothing by heating the device close to the melting point of \LN has been attempted \cite{LNReflow} with limited success (Q$\sim$3 $\times$ 10$^4$). In \cite{Creol}, the optical \Q is improved to 7.2 $\times$ 10$^4$ by etching a deposited cladding layer of slab waveguides with the drawback of reduced optical confinement. Recently, our group has demonstrated a \LN thin-film fabrication technology for manufacturing free-standing \LN RF MEMS devices \cite{MEMS13}, where we were able to achieve smooth and nearly vertical etching profile of \LN using a ion mill etching process on a full 4 inch 1um thick \LN thin-film on \LN substrate wafer. In contrast to focused ion beam etching, our process is batch fabrication compatible. Leveraging these MEMS fabrication techniques, we demonstrate here free-standing thin-film \LN microdisk resonator on silicon substrate with more than 50x better optical \Q. Moreover, the free-standing structure allows the coupling between the mechanical domain and optical domain, with which we also demonstrate acousto-optic modulation from the resonator.

\begin{figure}[h]
\centering
	\includegraphics[width=5in]{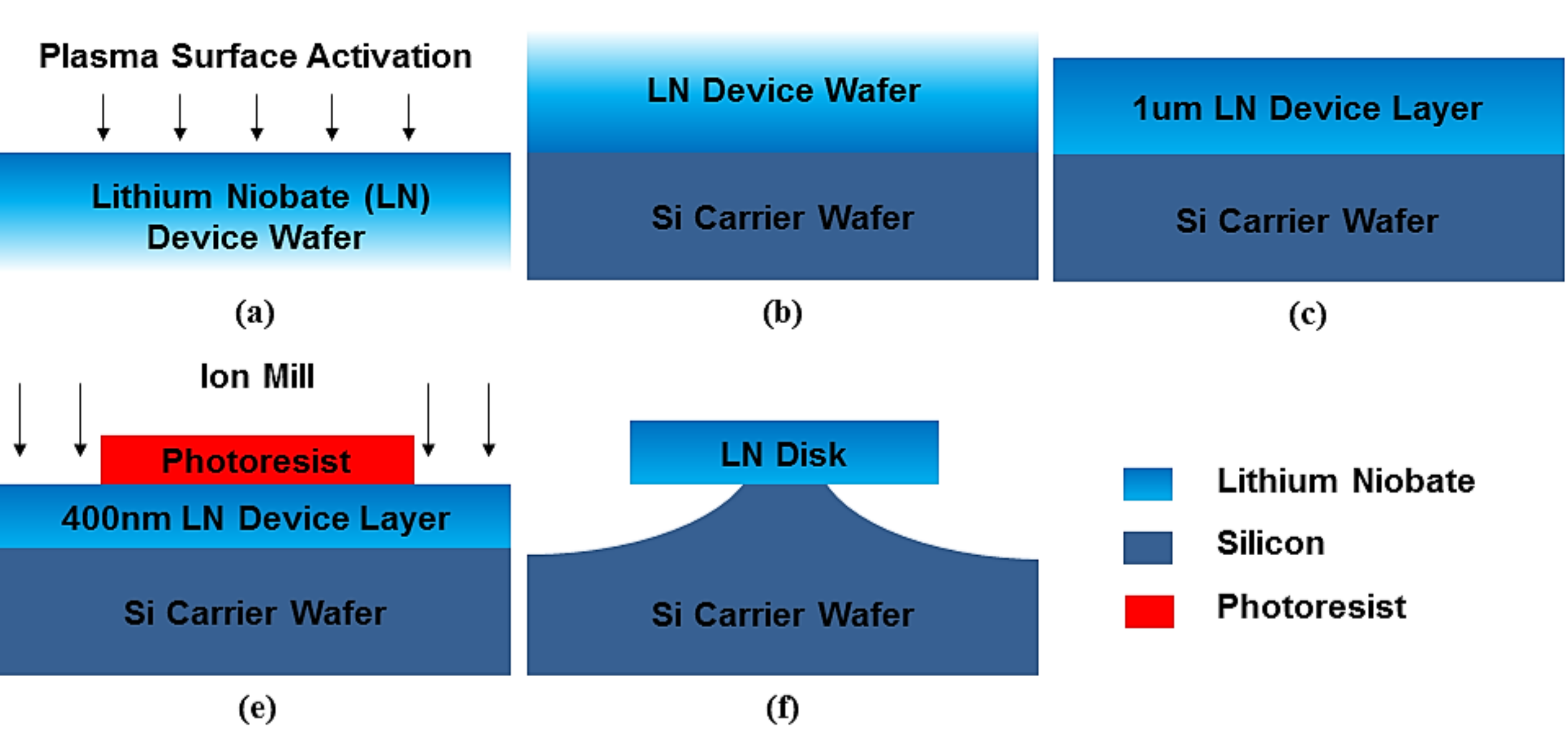}
	\caption{Fabrication process of the LN disk optical resonator: (a) Prepare the device wafer for bonding by plasma surface activation; (b) Direct bonding of the LN device wafer to Si carrier wafer; (c) Grounding the device wafer to 1um thickness; (d) Ion mill with photoresist mask to define device geometry; (e) XeF2 timed-etch release.}

\label{Fig:ProcFlow}
\end{figure}

\begin{figure}[h]
\centering
	\includegraphics[width=4in]{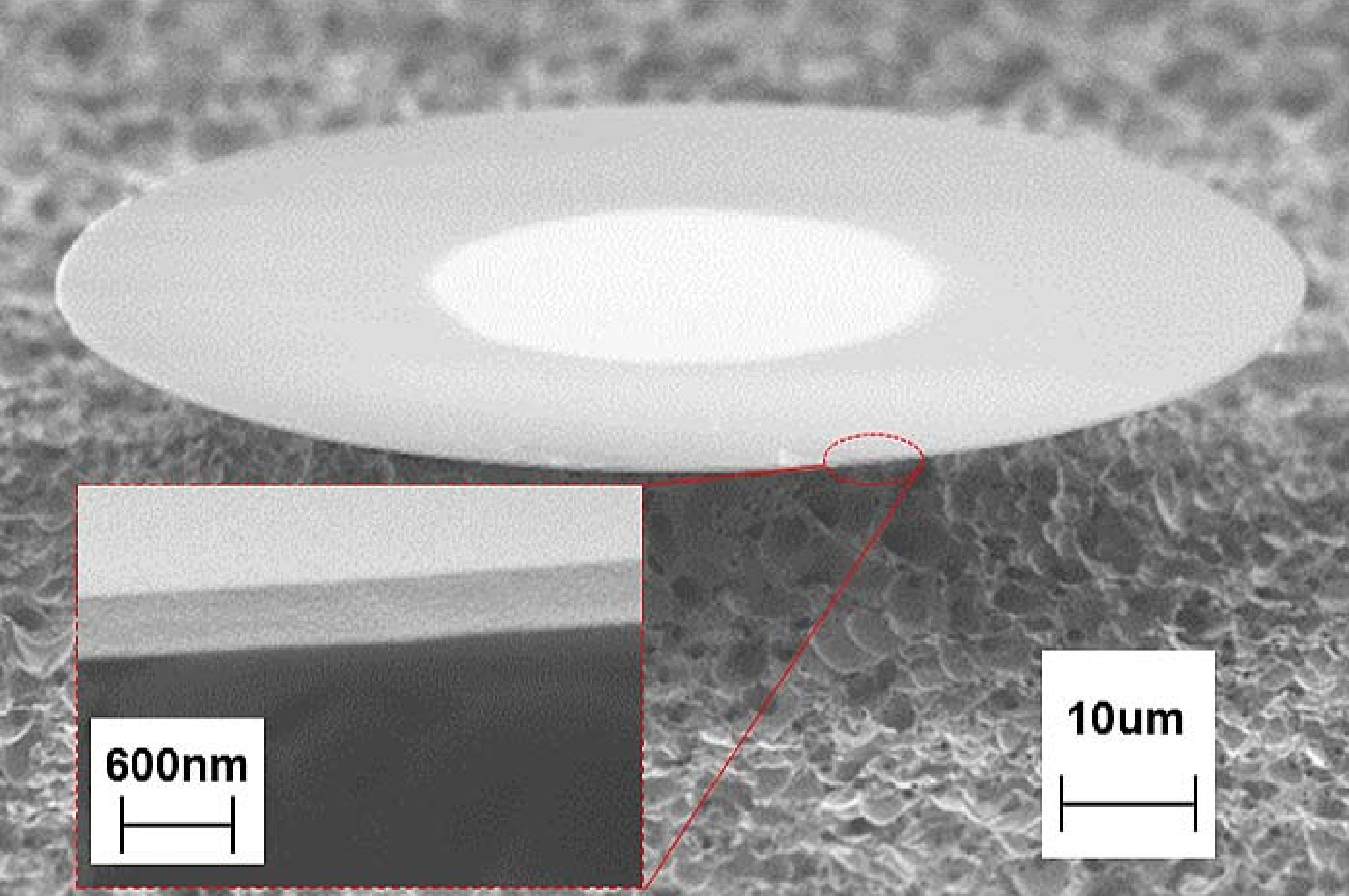}
	\caption{SEM of a 40um radius \LN photonic disk resonator; Inset: Zoom-in view of the rim showing smooth side wall, which is crucial for high optical \Q.}

\label{Fig:SmallDisk}
\end{figure}  

\section{Fabrication}

\begin{figure}[h]
\centering
	\includegraphics[width=5in]{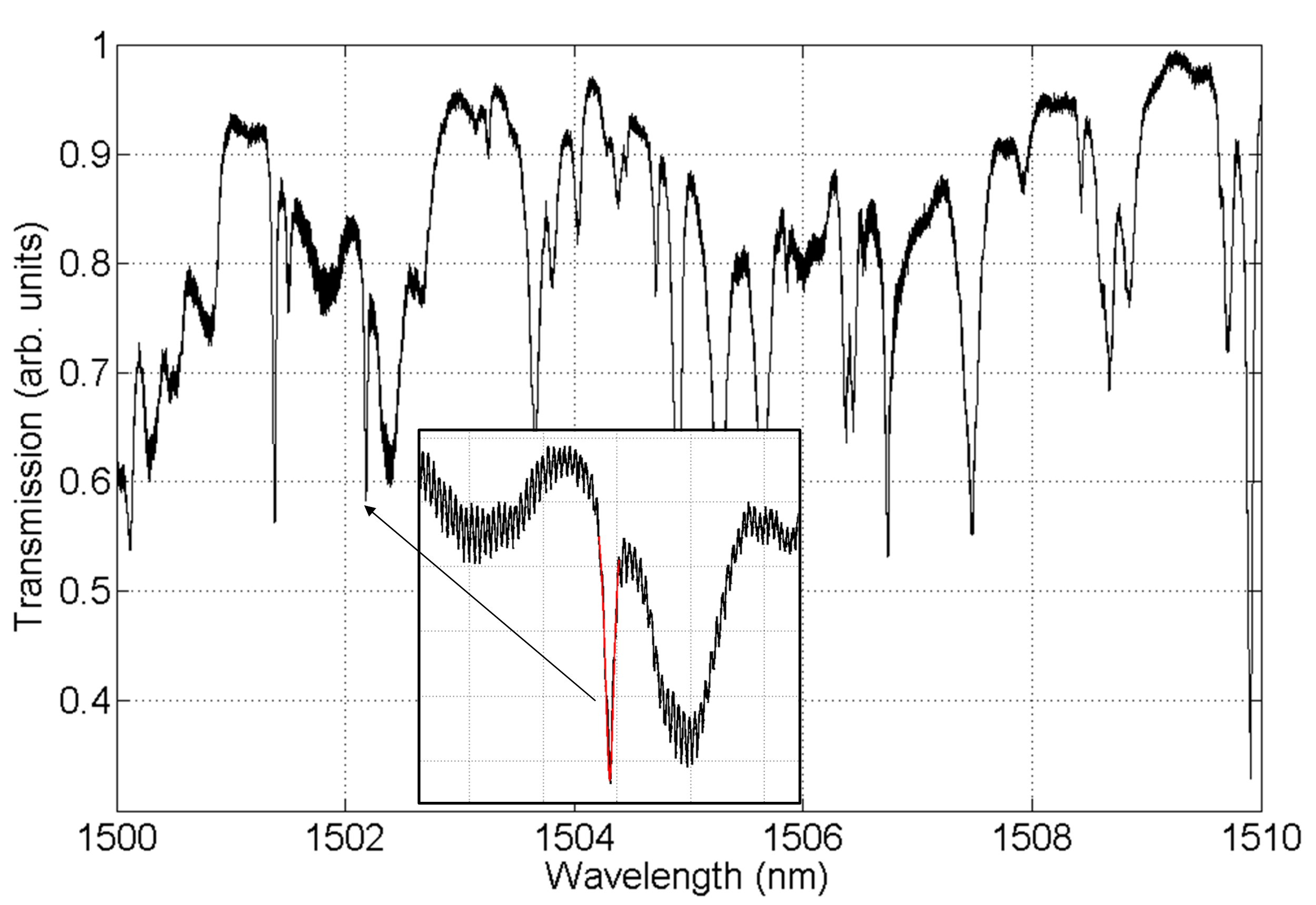}
	\caption{Optical transmission spectrum of a 35um radius resonator, and curve fitting of one peak to a Lorentzen.}
\centering
\label{Fig:QFit}
\end{figure}

There are three distinct fabrication challenges to fabricate the LN micro-photonic resonator on silicon substrate: 1. Achieving \LN thin-film with bulk single crystal \LN quality, 2. Achieving vertical side-wall and smooth surface profiles without mask residue, 3. Achieving clearance between the LN device and silicon substrate to enable efficient mode confinement.

The common technique for fabricating \LN thin-film is by ion-slicing \cite{GunterLN}. The drawback is that this process requires an annealing step to recover the crystal quality from the damage caused by the ion implantation. The annealing process unavoidably cause thermal stress in the film, which limits the yield for achieving \LN thin-film on full wafer scale. In contrast, we use a wafer bonding and grinding process as shown in Fig. \ref{Fig:ProcFlow}. We start with a Z-cut white \LN wafer. The bonding surface is activated by plasma (similar as in \cite{PlasmaActivation}). Then, the device wafer is flip-bonded to the Si handle wafer at room temperature. Therefore, the thermal stress is minimized, which allows us to achieve \LN thin-film on full 4in wafer. The device wafer is then ground down to 1um thickness. 

To date, etching of \LN have used either metal or silicon dioxide as a hard-mask, which increases the complexity and cause compatibility issues of the fabrication. In addition, any residue of metal left behind can interact with circulating photons causing optical absorption and scattering, which will significantly reduce the optical Q of the device. Here, we using 4um thick photo-resist as the mask with ion mill etching. The resist can then be easily rinsed off by acetone with sonication. If necessary, an additional O$_2$ plasma clean ensures that no residue is left behind. The Argon ion mill etching can be controlled to be self-balancing between etching and re-sputtering by carefully adjusting the ion beam incident angle, where the re-sputtering can act as an extra sidewall “protection” during the etching. With this process, we achieved sidewall angle of 87 degrees with $<$10nm surface roughness (inset of Fig. \ref{Fig:SmallDisk}), thus enabling low scattering loww and meeting the requirement of a smooth vertical sidewall that is necessary for future integrated waveguide coupling.

Silicon has higher refractive index ($\sim$3.4) than LN ($\sim$2.3). So any light that couples into the niobate disk will leak into the silicon substrate. Using XeF$_2$ dry-etch to undercut the niobate disk resonator (Fig. \ref{Fig:SmallDisk}) enables us to achieve $>$40um clearance between the niobate disk and silicon substrate, thereby enabling outstanding optical mode confinement in niobate and thus high optical Q. In addition, the free-standing disk structure will enable optomechanical \cite{Kippenberg} interactions between the mechanical and photonic modes.

\subsection{Optical characterization}

We use a tapered optical fiber \cite{Taper} to couple light from tunable near-IR laser (Santec TSL-510) to the niobate disk resonator (Fig. \ref{Fig:ProbeSetup}). A polarization controller is introduced to carefully adjust the light polarization in fiber such that it couples strongly with the photonic resonator. The transmitted optical signal is sent to a high-speed photodiode (Newport 1544A) and monitored on an Agilent (DSO9404A) oscilloscope. By sweeping the wavelength of the input light, we measure the optical transmission spectrum and extract the optical parameters (quality factor, group index, optical propagation loss) by fitting the transmission dip to a Lorentzian \cite{Taper}. Fig. \ref{Fig:QFit} shows the broad-band transmission spectrum of a 35um radius LN disk resonator from 1500nm to 1510nm. During the scan, the optical input power is kept below 20uW to ensure the resonator operates within the linear regime verified by the symmetric peaks. One resonance at 1502.61nm is fitted to the Lorentzen, and the inset shows the zoom-in view (red curve). The extracted intrinsic optical \Q is 484k, with a 95\% confidence bounds from 458K to 509K. The loaded \Q of the resonance is 39k, and the extrinsic \Q is 42k. The intrinsic \Q is $\sim$40 times better comparing to previous state of art.

\section{Acousto-optic modulation through optomechanical interaction }

\begin{figure}[ht]
\centering
	\includegraphics[width=5in]{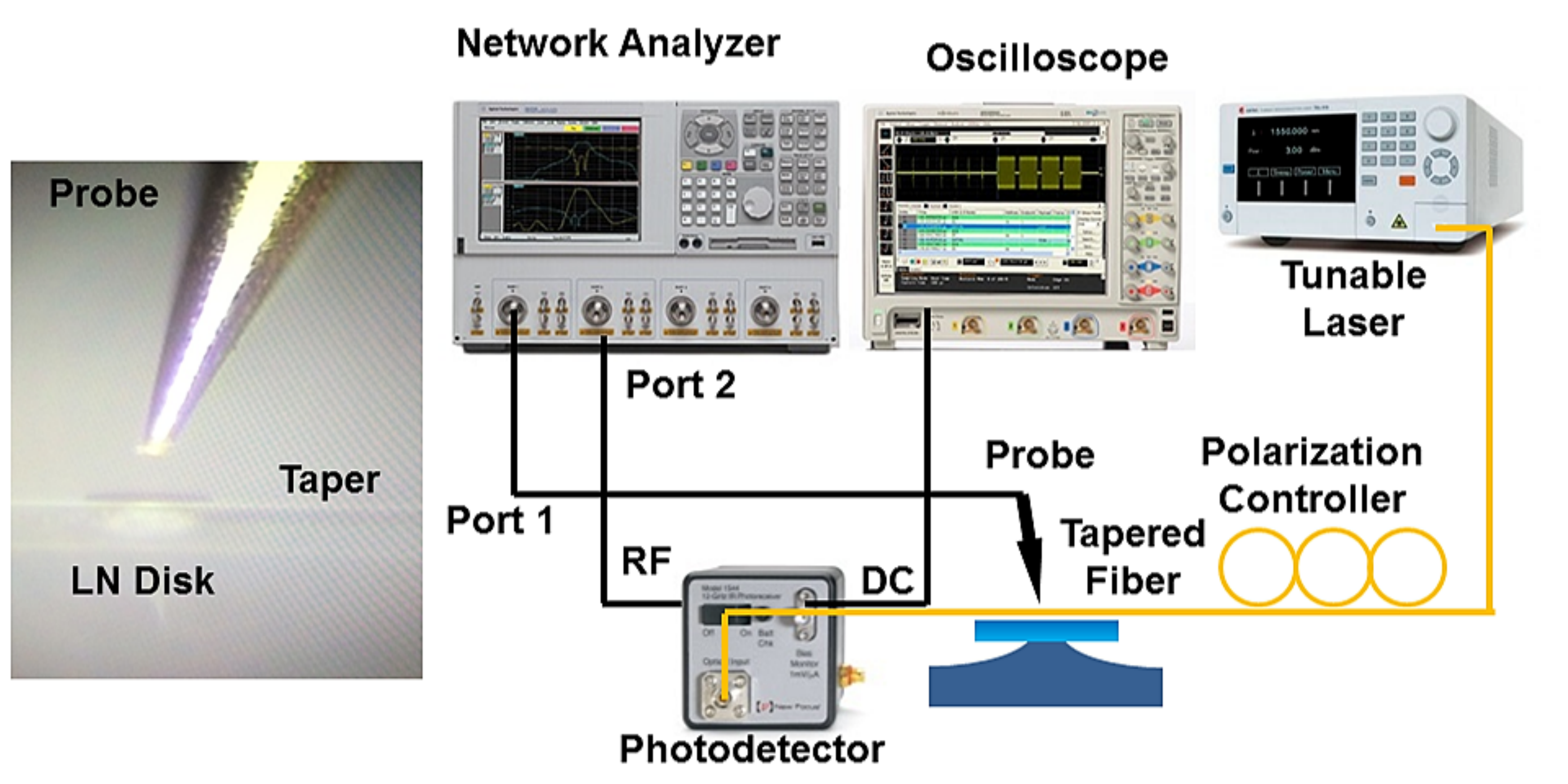}
	\caption{Side view of the acousto-optic modulation setup with a probe, and experimental setup of using probe to excite mechanical motion of the disk.}

\label{Fig:ProbeSetup}
\end{figure}

\begin{figure}[ht]
\centering
	\includegraphics[width=4.5in]{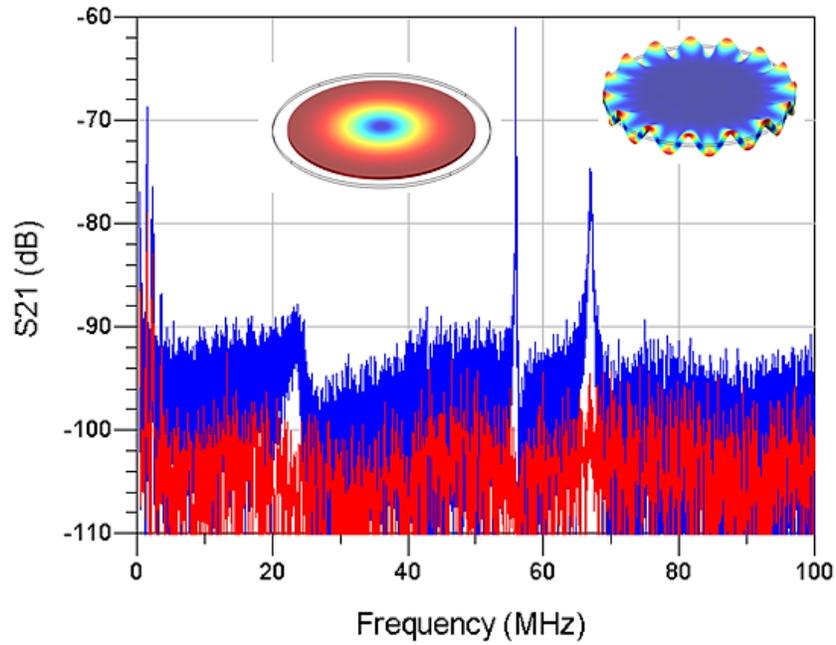}
	\caption{Scattering parameter measurement with the probe excitation.}

\label{Fig:G2S21}
\end{figure}

As the free-standing resonator body supports both mechanical and optical resonance, the optomechanical interaction can cause acousto-optic modulation of the transmitted light. The mechanical motion of the disk is excited by the gradient force of electric field applied by a probe (Fig. \ref{Fig:ProbeSetup}). This causes effective optical path length change causing the shift of the optical resonant frequency. When the wavelength of the input optical wave is biased at close to the 3dB transmission of the resonance, the frequency shift causes modulation of the transmitted optical power. Fig. \ref{Fig:ProbeSetup} shows the schematic of the experimental setup. The probe is excited by supplying an AC voltage from port 1 of an Agilent (N5230A) network analyzer. The RF output of the photodetector is connected to port 2 of the network analyzer, while the DC output is monitored on the oscilloscope to track the transmitted DC optical power. The output power from port 1 of the VNA is 8dBm, and Fig. \ref{Fig:G2S21} shows the scattering parameter measurement result from 300kHz to 100MHz. The red curve shows the s-parameters noise floor without coupling the taper to the resonator, while the blue curve shows the s-parameters when the taper is coupled to the resonator and the input wavelength is biased at a high \Q resonance around 1540nm. The wavelength is slowly tuned into the resonance from the short wavelength, and biased at the side of the resonance when the peaks in the $S_{21}$ curve is maximized. The transmitted DC optical power is 70uW. The $S_{21}$ shows two peaks around 50MHz with very good signal to noise ratio. The peak  at 56MHz in the $S_{21}$ measurement corresponds to the radial breathing mechanical mode as shown in the figure, while the 67MHz peak corresponds a shear mechanical wave traveling around the rim of the disk. The mechanical \Q of the radial mode is 2700 by measuring the 3dB bandwidth of the peak.

\section{Conclusion}

The MEMS-based fabrication technology presented in this work opens up new avenues to realize optical resonators, modulators, frequency doublers and frequency combs that leverage the multi-domain {RF, photonic, optomechanical} coupling in a monolithic LN-on-Silicon platform. We demonstrated free-standing thin-film \LN microdisk resonator with a intrinsic optical \Q of 4.84 $\times$ 10$^5$. In addition, we demonstrated acousto-optic modulation leveraging optomechanical interaction of the free-standing structure.

\end{document}